\documentclass[preprint]{revtex4}
\usepackage{epsfig}
\usepackage{graphicx}
\usepackage{dcolumn}
\usepackage{bm}
\def\la{\lower.5ex\hbox{$\; \buildrel < \over \sim \;$}}
\def\ga{\lower.5ex\hbox{$\; \buildrel > \over \sim \;$}}
\def\apj{ApJ}

\def\apjl{ApJL}

\def\aj{AJ}

\def\physrep{Physics Reports}

\begin{document}


\title{High Redshift Intergalactic Medium: Probes and Physical Models}
\author{Shiv K. Sethi}\email{sethi@rri.res.in}
\affiliation{Raman Research Institute, Bangalore 560080, India}


\begin{abstract}
Recent years have seen major advances in understanding  the state of the 
intergalactic medium (IGM) at high redshift. Some aspects of this understanding
are reviewed  here. In particular, we discuss: (1) Different 
probes of IGM like Gunn-Peterson test, CMBR anisotropies, and neutral hydrogen
emission from reionization, and (2) 
some models of reionization of the universe. 
\end{abstract}

\maketitle

\section{Indroduction} 
One of the outstanding issues in cosmology is to understand the reionization
of the universe. Following recombination of primordial plasma
at redshift $z \simeq 1000$, the universe is mostly neutral with
an ionization fraction $\simeq 10^{-4}$ (see e.g. \cite{peebles1},
\cite{padmanabhan1,padmanabhan2}). The Jean's mass at recombination
is $\simeq 10^6 \, \rm M_\odot$. At  $z \la 100$, the plasma 
thermally decouples from CMBR and its temperature decreases adiabatically,
$T_m \simeq 1/a^2$, which leads to a further decrease in Jean's mass.
During this 'dark and cold age' 
the density perturbations at scales above the Jean's
scale can grow. Figure~1 shows the ionization and thermal history 
of the universe along with the evolution of Jeans' mass. This age comes to an 
end when the first structures can collapse 
and the light from these objects can reionize and reheat the universe
(for a recent review see \cite{barkana} and references therein) . 
Therefore the epoch of re-ionization holds important clues about the 
way first structures formed and can potentially distinguish between
different models of structure formation. In the standard $\Lambda\rm CDM$
models the first structures to collapse would be just above the Jean's 
length (see e.g. \cite{padmanabhan1}). 
Another  crucial question in this regard is whether these
structures could cool fast enough to form stars. Many of these 
issues will be discussed in this review.

Different ongoing and potential probes of intergalactic medium can
reveal the nature of the re-ionization of the universe. One of the 
most important and the oldest is the Gunn-Peterson test (see e.g. \cite{peebles1} and reference therein), which is 
very sensitive to the neutral fraction in the intergalactic medium. 
CMBR anisotropy measurements are another powerful and complementary 
probe, as they are sensitive to the ionized component of the intergalactic 
medium (see \cite{hu1}, \cite{bond} and references therein). In future, 
it might be possible to directly observe the first sources that 
re-ionize the universe.
In addition, the transition from neutral to ionized universe might also
be detected in neutral hydrogen emission (see e.g. \cite{madau}, \cite{shaver}, \cite{tozzi}).  

To sum up the observational status: Recent detection of 
temperature-polarization cross-correlation in CMBR suggests that the 
redshift of reionization
$z_{\rm reion} \simeq 17\pm 4$ \cite{kogut}. 
Gunn-Peterson probes suggest that 
the universe is highly ionized upto $z \simeq 5$, but might be making
a transition from highly ionized to neutral for $5 \la z \la 6$ (\cite{fan}, \cite{djorgovski}, \cite{becker}) . 
These two observations together throw open the interesting possibility
that the universe went through two phases of re-ionization.

This article is divided  into two parts. In the first part, probes that
give a clue about the reionization epoch will be discussed. In the 
second part, we will discuss the nature of ionizing  sources. Throughout
this review we
use the currently-favoured FRW  model: spatially flat
with $\Omega_m = 0.3$ and $\Omega_\Lambda = 0.7$ (\cite{spergel}, \cite{perlmutter},
\cite{riess}) with  $\Omega_b h^2 = 0.02$ \cite{spergel}, \cite{tytler}) and
$h = 0.7$ \cite{freedman}.

\section{Probes of Ionization at high redshifts}

\subsection{Gunn-Peterson effect}
High redshift sources should show absorption at frequencies close to 
Lyman-$\alpha$ (1216~$\rm \AA$) line owing to scattering from IGM 
neutral hydrogen (HI). 
This test applied to high redshift quasars which generally
have a strong Lyman-$\alpha$ emission line means that the blueward  side 
of the Lyman-$\alpha$ line should show strong absorption as compared
to the redward side (Gunn-Peterson (GP) test).  
The optical depth to the Lyman-$\alpha$ scattering
from  HI can be calculated (see e.g. \cite{peebles1}):
\begin{equation}
\tau_{\rm \scriptscriptstyle GP}(\nu_0) \simeq  4 \times 10^{5} x_{\rm \scriptscriptstyle HI} \left (h \over 0.7 \right ) \left (\Omega_b \over  0.045 \right )  \left (0.3 \over  \Omega_m \right )^{1/2}  \left (1+z \over  6 \right )^{1.5}
\label{eq:gpt}
\end{equation}
Here $x_{\rm \scriptscriptstyle HI} \equiv  n_{\rm \scriptscriptstyle HI}/n_{\rm \scriptscriptstyle H}$ ($n_{\rm \scriptscriptstyle H} \simeq 0.92 n_b$) is 
the neutral fraction of hydrogen and  $\nu_0 = \nu_\alpha/(1+z)$. Two points 
worth noting in Eq.~(\ref{eq:gpt}) are: (1) Owing to the resonant nature of 
Lyman-$\alpha$ scattering, optical depth at observed frequency $\nu_0$ 
gives direct information about the neutral fraction at redshift $(1+z) = \nu_\alpha/\nu_0$, and (2) more importantly, this test is extremely sensitive to 
the neutral fraction of hydrogen, even a neutral fraction as small as a part 
in hundred thousand can fully absorb light shortward of Lyman-$\alpha$ in the 
quasar spectrum.

Since 1960s when the first high redshift quasars whose Lyman-$\alpha$
emission  could be 
detected from ground-based telescopes were discovered, GP test has been 
applied to study the ionization of the universe at high redshifts. Till 2000,
none of the observed quasars upto a redshift  $\simeq 5$ showed any 
GP absorption, which means the universe is ionized to better than a part in
a million upto $z \simeq 5$. The discovery of several quasars at redshifts 
above  six  by SDSS survey made it possible to apply GP tests at even higher 
redshifts. GP absorption has been  detected 
in several quasars at redshifts $z \ga 5.7$ (\cite{becker}, \cite{djorgovski}, \cite{fan})  . Observations suggest that
the neutral fraction increases rapidly between 
$z \simeq 5.5$ to $z \simeq 6.2$ (see e.g. \cite{fan}). 
These observations might mean that
the universe is becoming ionized owing to the formation of the first 
structures at $z \simeq 6$. However a straightforward interpretation of these
results is not easy. Even for the best quasar spectrum and using the 
Lyman-$\beta$ line, which has a smaller oscillator strength, the 
GP optical depth  $\tau_{\rm \scriptscriptstyle GP} \ga 25$ 
(\cite{white}). Using Eq.~(\ref{eq:gpt}), this implies that
$x_{\rm \scriptscriptstyle HI} \ga 10^{-4}$ i.e. the  universe can be 
almost fully ionized. Use of semi-analytic models which take into account the 
clumpiness of the IGM give more stringent constraints  on the 
neutral fraction and  give  $x_{\rm \scriptscriptstyle HI} \ga 10^{-3}$
at $z \simeq 6$ \cite{fan}.  Even though the observations are unable
to conclude that the universe made a transition from fully ionized to
almost fully neutral between redshift of five and six, the ionized fraction
certainly evolves very rapidly in this redshift range, 
and this can have important
implications for the models of structure formation. 

\begin{figure}
\epsfig{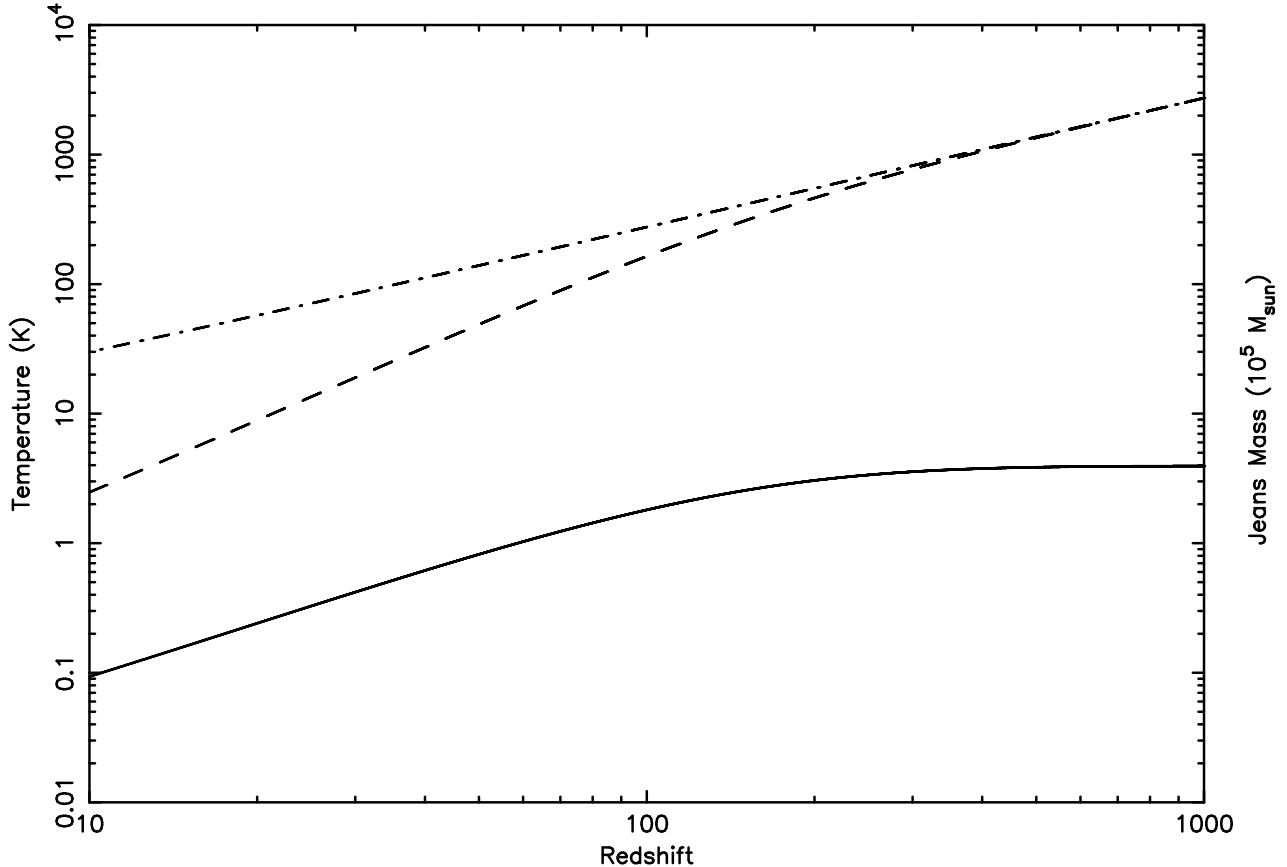}
\caption{The temperatures of CMBR and matter are plotted alongside 
the evolution of Jeans mass. The solid curve is Jeans mass in units of 
$10^5 \, \rm M_\odot$. The dashed and dot-dashed curves correspond to matter 
and CMBR temperature (in Kelvin), respectively.}
\label{fig:f1}
\end{figure}

\subsection{CMBR anisotropies}
CMBR anisotropies provide a complementary approach to the ionization history
of the universe as compared to the GP test, as they are sensitive to the 
ionized component of the universe. The physics of CMBR temperature 
anisotropies at the last scattering surface 
 and various data analysis issue are covered elsewhere in this 
volume (Subramanian, this volume).
 Here we shall discuss the implications of 
reionization on the CMBR temperature anisotropies. To highlight the 
effect of reionization on the polarization anisotropies we discuss more 
fully the CMBR polarization anisotropies. 

The universe recombines at $z \simeq 1000$. Following recombination the 
ionized fraction ($x_e \equiv n_e/n_{\rm \scriptscriptstyle H}$)
 in the universe is   $\simeq 10^{-4}$ 
for $z \simeq 100$ (see e.g. \cite{peebles1}).  The mean free path of the 
CMBR photons to Thompson scattering 
 exceeds the local Hubble radius in the post-recombination era and therefore
the universe is 'transparent' to the CMBR photons. Following reionization, 
the ionized fraction might reach nearly unity and a small fraction of 
CMBR photons might re-scatter again. An important quantity in studying 
CMBR quantity is visibility function which is 
the normalized probability that the photon scattered 
in a range  $z$ and $z +dz$. It is defined as:
 $V(\eta_0,\eta) = \dot \tau \exp(-\tau)$, here $d\eta = dt/a$ is the conformal time
and $\tau = \int_{\eta_0}^{0} x_e n_{\rm \scriptscriptstyle H} a  \sigma_t cd\eta$. 
In Figure~2 we show the visibility function for two models. For the 
model with early reionization, the visibility function get important 
contribution from redshift of reionization. One can define the optical
depth to the reionization surface: $\tau_{\rm reion} =  \int_0^{z_{\rm reion}} x_e n_{\rm \scriptscriptstyle H}  \sigma_t cdt$; this is the 
fraction of photons that re-scattered in the reionized universe for 
$\tau_{\rm reion} < 1$. For 
$z_{\rm reion} \simeq 6$, $\tau_{\rm reion} \simeq 0.05$ and 
$\tau_{\rm reion} \propto (1+z_{\rm reion})^{3/2}$. 

\begin{figure}
\epsfig{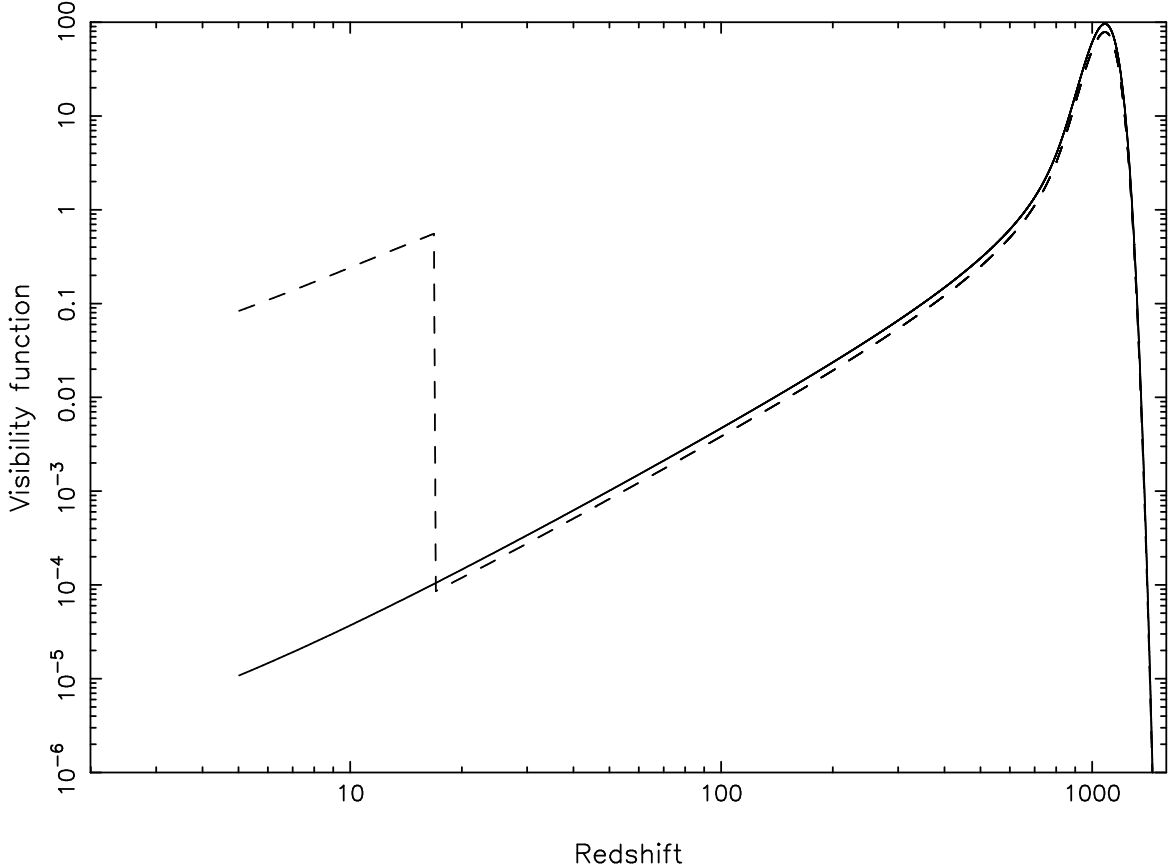}
\caption{Visibility function, defined as
$d\tau/d\eta \exp(-\tau) H_0^{-1}$, is plotted for different models. The
solid and the dashed  curves are for the standard recombination
and a model in which
the universe reionizes at $z =17$, respectively.}
\label{fig:f2}
\end{figure}

The temperature and polarization anisotropies can be computed by solving the 
Boltzmann equations for the photon distribution function. Equations 
appropriate for studying the effect of reionization for scalar 
perturbations, for a given
wavenumber ${\bf k}$ and line of sight ${\bf n}$, are (see e.g. \cite{hu1},
Zalddariaga 1997, \cite{bond},\cite{dodelson}):
\begin{eqnarray}
\dot \Delta_T + ik\mu \Delta_T & \simeq &  \dot \tau (\mu v - \Delta_T) \nonumber \\
\dot \Delta_P + ik\mu \Delta_P &= & \dot \tau (\Pi(\mu)[\Delta_{T2} +\Delta_{P2}- \Delta_{P0}] -\Delta_P) 
\label{eq:tpeq}
\end{eqnarray}
Here $\mu = {\bf k.n}$, $\dot \Delta_T {} \equiv \partial \Delta_T/\partial \eta$, $ \Delta_T \equiv \Delta T/T(k,\mu,\eta)$. The polarization 
anisotropies $\Delta_P \equiv \Delta P/T(k,\mu,\eta)$ are 
  in the plane perpendicular to the ${\bf k}$ vector, which 
is taken to be  parallel to the z-axis, to exploit the axial symmetry of the 
problem.  In this case, only one Stokes parameter $Q$ is  non-zero  and 
$\Delta P = Q T$.  Also $\dot \tau = n_b x_e \sigma_T a$,
 $\Pi(\mu) = 0.5(1-P_2(\mu))$. $v$ is the electron velocity; we also assume 
 curl-free  velocity fields which allows us to express:
 ${\bf v.n} = {\bf v.k}$. The angular moments of temperature and 
polarization anisotropies are: $\Delta_{\ell} = {1\over 2} \int_{-1}^1 d\mu P_\ell(\mu) \Delta(k,\mu)$. Even though the temperature equation is 
only appropriate for studying the effect of reionization on 
anisotropies generated at last scattering surface, the polarization equation is exact and 
can also be used to study the generation of perturbations at the last scattering surface.

\begin{figure}
\epsfig{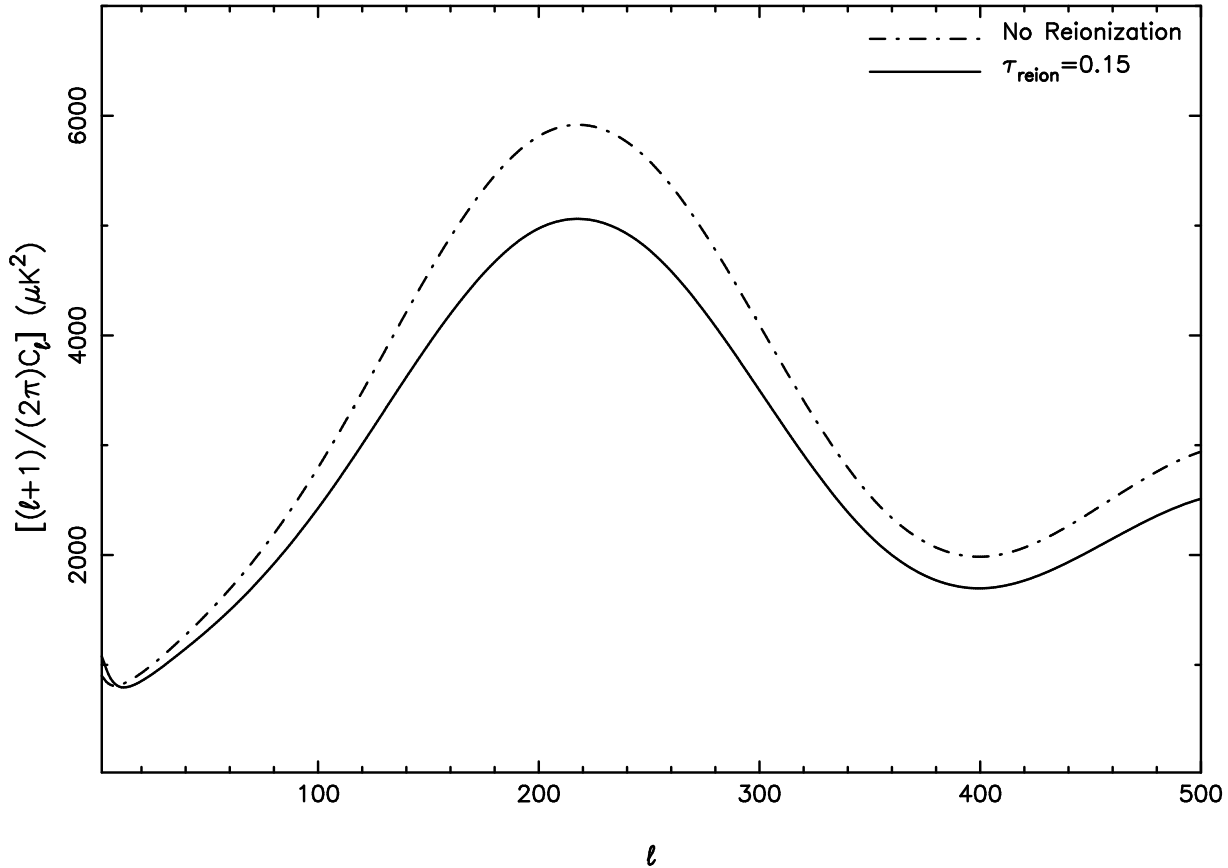}
\caption{The effect of reionization CMBR temperature anisotropies 
is shown. The main effect of reionization on temperature anisotropies
is to generate secondary signal for $\ell \la 20$ and suppress primary 
anisotropies by a factor $\exp(-2\tau_{\rm reion})$ for larger $\ell$. } 
\label{fig:f3}
\end{figure}

To understand the essential physics of CMBR anisotropies in reionized models, we only 
solve for anisotropy evolution for  one wavenumber ${\bf k}$. The quantity measured 
by experiments is the two point function of this quantity summed over all the wave-numbers
(for details see articles by  Subramanian, this volume). 

We begin by studying the effect of reionization on the temperature 
anisotropies. Eq~(\ref{eq:tpeq}) can be solved to give:
\begin{equation}
\Delta_T(\eta) = \Delta_T(\eta_{\rm rec})\exp(ik\mu(\eta_{\rm rec}- \eta))\exp(-\tau(\eta_{\rm rec},\eta)) + \mu \int d\eta'v(\eta')V(\eta,\eta') \exp(ik\mu(\eta'- \eta))
\label{eq:tesol}
\end{equation}
The first term in the solution means the anisotropies generated at the 
last scattering surface are exponentially damped by reionization (the 
solution is only correct for scales smaller than the size of local horizon,
i.e. $k \ga \eta^{-1}(z_{\rm reion})$. It is not reflected in the 
solution  owing to dropping a term $\propto
\Delta_{T0}$ in the temperature equation 
(for details see \cite{hu3})). For example, if the universe reionized at $z \simeq 50$ which gives $\tau_{\rm reion} \simeq 1$ .i.e. 
all the photons from 
the last scattering surface are re-scattered following reionizatio,
 this means that all anisotropies at scales smaller than the angular
scale correspond to $\ell \simeq  \eta_0/\eta(z_{\rm reion}) \simeq 10$ 
are wiped out. As we shall see the second term in the solution doesn't 
contribute much to the anisotropies at small scales either. This means
that for a reionization redshift $\simeq 50$  no anisotropies should be 
observed for $\ell \ga 10$, which is in direct contradiction with 
observations (e.g. WMAP observations detect anisotropies for $\ell \simeq 600$). Therefore 
the redshift of reionization should be small enough such that
only a small fraction of CMBR photons are re-scattered.  To compute the 
second term in Eq.~(\ref{eq:tesol}), we can assume the visibility to 
be a normalized Gaussian with a width $\Delta\eta_{\rm reion}$, this gives
\begin{equation}
\Delta_T \propto \mu \tau_{\rm reion} v \exp\left[-(k\mu \Delta\eta_{\rm reion}/2)^2\right ]
\end{equation}
This shows that CMBR anisotropies generated during the epoch of 
reionization owing to Doppler scattering off electrons are suppressed for $k \ga 1/\Delta\eta_{\rm reion}$. The generated signal as expected is $\propto \tau_{\rm reion}$   For realistic 
ionization histories $\Delta\eta_{\rm reion} \simeq \eta_{\rm reion}$. 
We show in Figure~3, the effect of reionization on the temperature
 anisotropies. The net effect of reionization on the CMBR temperature anisotropies can
be summarized as: (a) anisotropies at small scales are suppressed exponentially
as $\exp(-2 \tau_{\rm reion})$, (b) anisotropies at scales corresponding to $\ell \la 10$ escape this exponential damping and also new anisotropies are
generated at these scales. As seen in Figure~3, reionization causes a relative
decrease in the small scale anisotropies. The minimum 
error in detecting the angular power spectrum of CMBR anisotropies 
at any $\ell$ is $\Delta C_\ell \simeq \sqrt{2/(2\ell+1)} C_\ell$ (Cosmic variance)
 and for all CMBR anisotropy experiments the error on $C_\ell$ is dominated by 
the cosmic variance for $\ell \la 300$. 
The reionization signal  is very difficult
to detect owing to uncertainty in the overall normalization of the CMBR 
anisotropies which is compounded by cosmic variance. From WMAP data 
all the information about the reionization comes from the polarization signal.

\begin{figure}
\epsfig{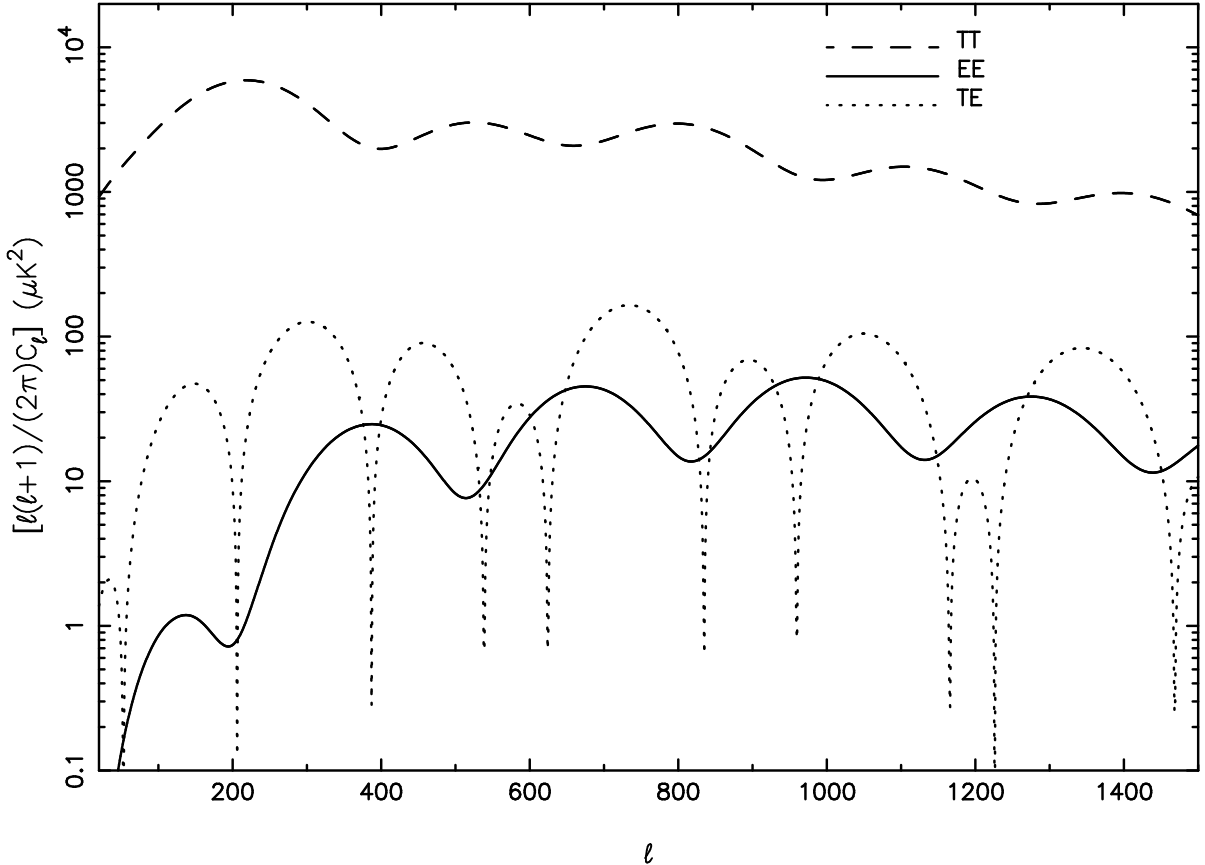}
\caption{Primary CMBR temperature (TT) and polarization (EE) anisotropies and their cross-correlation (TE)  are shown; the absolute value of the 
temperature-polarization cross-correlation is plotted. } 
\label{fig:f4}
\end{figure}

In second order in perturbation theory, reionization causes potentially detectable 
anisotropies for $\ell \ga 1000$; for $\ell \ga 2000\hbox{--}3000$ this signal can dominate
the signal from primary anisotropies. The most important second order contribution to
temperature anisotropies comes from Vishniac 
effect (for details see \cite{vishniac}, also see  \cite{hu1},
\cite{dodelson}, \cite{hu3} for host of other second order effects). 

From Eq.~(\ref{eq:tpeq}), it can be seen that the source of polarization anisotropy is 
the quadrupole of the temperature anisotropy. The generation  of temperature and 
polarization anisotropies  at the last scattering surface is generally 
an involved process  and  Eqs.(\ref{eq:tpeq}) have  to be solved  numerically (see
e.g. \cite{seljak}). However, several relevant  assumptions allows one to get
an analytic  insight into the problem (\cite{hu2}, \cite{zaldarriaga2}). 
First simplification occurs because the photon-baryon plasma can be treated as tightly-coupled
for most of the period during which the recombination lasts. The tight coupling approximation is valid for scales corresponding to $k \la l_f^{-1}$, where $l_f$ is the (comoving) 
mean free path
of the photons for Thompson scattering; $l_f \simeq 1 \, \rm Mpc$ at the last scattering
surface for a fully ionized universe, which means that tight coupling approximation is valid
for  most scales of interest. (The scale of interest for tight-coupling in not 
$l_f$ but the scale of photon diffusion at recombination which for a fully ionized plasma
is roughly 10 times larger than $l_f$; free-streaming of photons at recombination further
increase this scale; for detailed discussions and implications of this for CMBR anisotropies 
see   \cite{hu2}). A second simplification, closely related to the first 
but physically distinct, occurs because the width of visibility function at the last scattering surface 
 corresponds to scales $\la 10 \, \rm Mpc$, and hence for studying physical 
processes at much larger scales  the recombination 
can be treated as instantaneous (for caveats see \cite{hu2}). Out interest here is 
in scales that are super-horizon at the last scattering surface
($k^{-1} \ga 100 \, \rm Mpc$, comoving),  corresponding to angular scales $\ell \la 200$. Therefore
we will use tightly coupled approximation and not discuss the effects of photon diffusion. 
Using these approximations, adiabatic initial conditions
 and also the fact that the ratio of baryon to 
photon energy density $\rho_b/\rho_\gamma \simeq 25 \Omega_b h^2 \ll 1$ at the epoch of decoupling for 
acceptable models of primordial nucleosynthesis, the temperature and polarization anisotropies
generated at the last scattering surface are:
(for details see \cite{hu2}, \cite{zaldarriaga2}):
\begin{eqnarray}
\Delta_T(k,{\bf n},\eta) &  = & {1 \over 3} \Phi({\bf k},\eta_{\rm rec})\cos(kc_s\eta_{\rm rec})\exp[ik\mu(\eta_{\rm rec} - \eta)] \nonumber \\
\Delta_P(k,{\bf n},\eta) &  = & 0.17(1- \mu^2)kc_s \Delta\eta_{\rm rec}\Phi({\bf k},\eta_{\rm rec}) \sin(kc_s\eta_{\rm rec})\exp[ik\mu(\eta_{\rm rec} - \eta)]
\label{tpeqsol}
\end{eqnarray}
Here $\Phi({\bf k},\eta_{\rm rec})$ is the Fourier component of the Newtonian potential 
at the last scattering surface; $\Delta\eta_{\rm rec} \simeq 10 \, \rm Mpc$ is the 
comoving width of LSS; $c_s \simeq 1/\sqrt{3}$ is the sound velocity in the coupled
photon-baryon fluid. These large scale solutions show that: (a) temperature 
and polarization anisotropies are correlated with each other, (b) the amplitude of 
polarization anisotropies is suppressed by a factor $kc_s \Delta\eta_{\rm rec}$; e.g. 
for $k \simeq 0.01 \, \rm Mpc^{-1}$ this factor is roughly $1/10$. This 
suppression is owing to the fact that polarization anisotropies $\propto \Delta_{T2}$. 
In the strict tight coupling approximation, only the temperature monopole and dipole are 
non-zero in the comoving frame. A small quadrupole is generated owing to 
free streaming of photons before they  scatter for the last time; and this quadrupole is 
suppressed with respect to the monopole and dipole,  which constitute the primary sources of temperature anisotropies, by the factor $\simeq kc_s \Delta\eta_{\rm rec}$. 
As we shall see below, 
in the reionized models this suppression is absent.  In Figure~4 we show the 
CMBR temperature and polarization anisotropies, generated at the last scattering surface. 

\begin{figure}
\epsfig{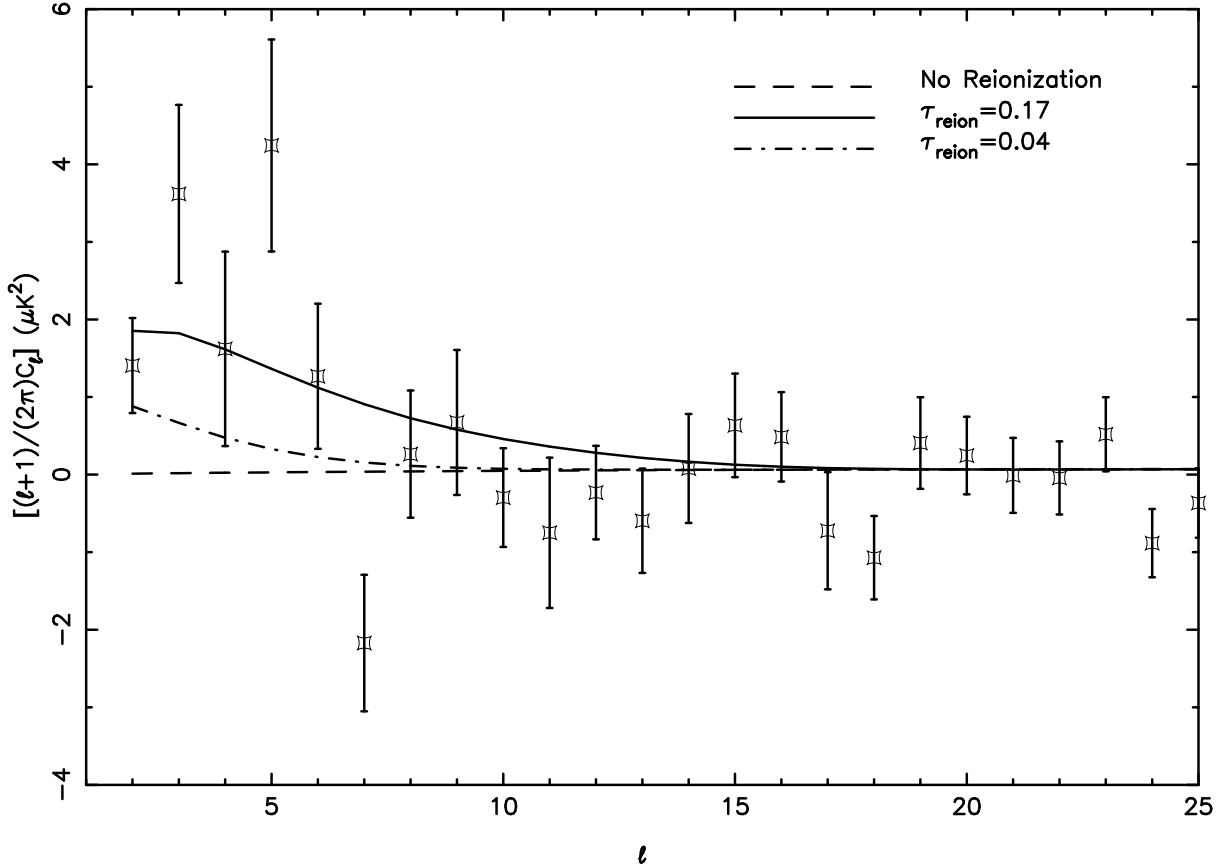}
\caption{The effect of reionization on the temperature-polarization
cross-correlation is shown. Also shown are the WMAP data points 
with one-sigma error bars \cite{kogut}} 
\label{fig:f5}
\end{figure}

In the reionized models, a fraction CMBR photons scatter again after the epoch of 
reionization. As argued above, this fraction is small as many of the features of 
the primary anisotropies have already been detected. We have already discussed the effect
of reionization on temperature anisotropies and argued that the reionization signal
is very difficult to discern from temperature anisotropies alone. Like temperature 
anisotropies, one of the effects of reionization would be to wipe out polarization anisotropies
generated at the last scattering surface for scales $\la \eta_{\rm reion}$. From 
Eq.~(\ref{eq:tpeq}), the generation of new polarization anisotropies will be proportional
to the value of $\Delta_{T2}$ at the epoch of reionization. Eq.~(\ref{eq:tpeq}) 
can simplified further by dropping terms of polarization
monopole and quadrupole of polarization anisotropies in the RHS of the equation, as these terms are negligible compared to the temperature quadrupole (see below). Eq.~(\ref{eq:tpeq}) 
then give the following solution for the generation of polarization anisotropies following 
reionization:   
\begin{equation}
\Delta_P = \int d\eta' \Pi(\mu) \Delta_{T2}(\eta')V(\eta,\eta')\exp(ik\mu(\eta'- \eta))
\end{equation}
Most of the contribution to the integral will come from  close to the 
reionization epoch (typically  $\simeq \Delta\eta_{\rm reion} \simeq \eta_{\rm reion}$,
 the width of visibility function if it is 
approximated  as a Gaussian as above). The amplitude of the contribution is 
proportional to the temperature quadrupole at the epoch of reionization. 
As argued above the 
temperature quadrupole is suppressed at the last scattering surface owing to 
tight coupling approximation that holds at that epoch. Following recombination the 
photons free stream which allows to temperature quadrupole to increase. From Eq.~(\ref{tpeqsol}), we can get the temperature quadrupole at the reionization epoch 
from taking the angular moment of the equation (for details see e.g. \cite{zaldarriaga1}):
\begin{equation}
\Delta_T^{(2)}(\eta_{\rm reion}) =  {1 \over 3} \Phi(k,\eta_{\rm rec})\cos(kc_s\eta_{\rm rec}) j_2[k(\eta_{\rm reion}-\eta_{\rm rec})]
\end{equation}
Using the fact that the maximum value of $j_2(x) \simeq 0.3$, and comparing with 
the polarization anisotropies generated at the last scattering surface,Eq.~(\ref{tpeqsol}),  
we can see that $\Delta_T^{(2)}(\eta_{\rm reion})$ doesn't suffer the $k$ dependent 
suppression, and 
is of the order of temperature anisotropies at the last scattering surface. At  $k \ll 0.1$,
it can be several orders of magnitude more than the polarization anisotropies generated at the last scattering surface. The polarization anisotropies at the reionization epoch 
peak at the characteristic scale $k \simeq 2/\eta_{\rm reion}$,
 which corresponds to 
an angular scale $l \simeq k \eta_0 \simeq 5\hbox{--}10$ for $z_{\rm reion} \simeq 15\hbox{--}50$.

In Figure~5 we show the effect of reionization on the temperature-polarization cross-
correlation power spectrum; also shown are the observations of WMAP. Figure~5 shows that
the enhancement of power at $\ell \le 10$ cannot be explained within the framework of 
no reionization models. Also shown are the predictions of a model 
in which the universe 
reionized at $z \simeq 5.5$, which the GP observations discussed above 
might be suggesting
and the best fit model to the WMAP observations. The best fit model 
requires $\tau_{\rm reion}\simeq 0.15$, which implies the epoch of 
reionization corresponds to  $z_{\rm reion} \simeq 15$.

\subsection{High Redshift HI}

Another possible probe of the reionization epoch is to observe  the 
neutral component 
of hydrogen thorough the epoch of reionization. The neutral fraction of 
hydrogen changes from near unity to zero during the epoch of reionization.
This change can potentially be observed using  the hyperfine transition
of the hydrogen atom at $\nu_\star = 1420 \, \rm MHz$. The quantity 
of interest here is the spin temperature
of hydrogen defined as:
\begin{equation}
{n_2 \over n_1} = 3 \exp(-T_\star/T_s)
\end{equation}
Here $n_2$ and $n_1$ are the populations of the hyperfine states.
$T_\star = h \nu_\star/k = 0.06 \, \rm K$. As the only radio source at the 
high redshift is CMBR, HI in hyperfine transition can seen against the 
 CMBR in emission or absorption. The observed quantity then is the 
deviation of CMBR from a black body at radio frequencies. 
The observed difference is:
\begin{equation}
\Delta T_{\rm CMBR} = -\tau_{\rm \scriptscriptstyle HI}(T_{\rm CMBR} -T_s)
\label{eqh1}
\end{equation}
Here $\tau_{\rm \scriptscriptstyle HI} = \sigma_\nu N_{\rm HI} T_\star/T_s$,
with the HI column density $N_{\rm HI} = \int n_{\rm \scriptscriptstyle HI} d\ell$; $\sigma_\nu = c^2 A_{21}\phi_\nu/(4\pi \nu_\star)$;  
 $A_{21} \simeq 1.8 \times 10^{-15} \, \rm sec^{-1}$ 
and $\phi_\nu$ is the line response function. The spin temperature 
 is determined from detailed balancing between various processes that
 can alter the 
relative populations of the two levels \cite{field1, field2}:
\begin{equation}
T_s = {T_{\rm cmbr} + y_c T_K + y_\alpha T_\alpha \over 1 +  y_c + y_\alpha}
\label{eqts}
\end{equation}
Here $y_c \propto n_{\rm \scriptscriptstyle H}$  and $y_\alpha \propto n_{\rm \scriptscriptstyle \alpha}$, with $n_\alpha$ being the number density of 
Lyman-$\alpha$ photons  correspond, respectively,  to   relative 
probabilities with which 
collisions between atoms and the presence of Lyman-$\alpha$ photons  determine
the level populations;  $T_K$ is the matter temperature. In the pre-reionization era, there are no Lyman-$\alpha$ photons, and therefore $y_\alpha = 0$. 

\begin{figure}
\epsfig{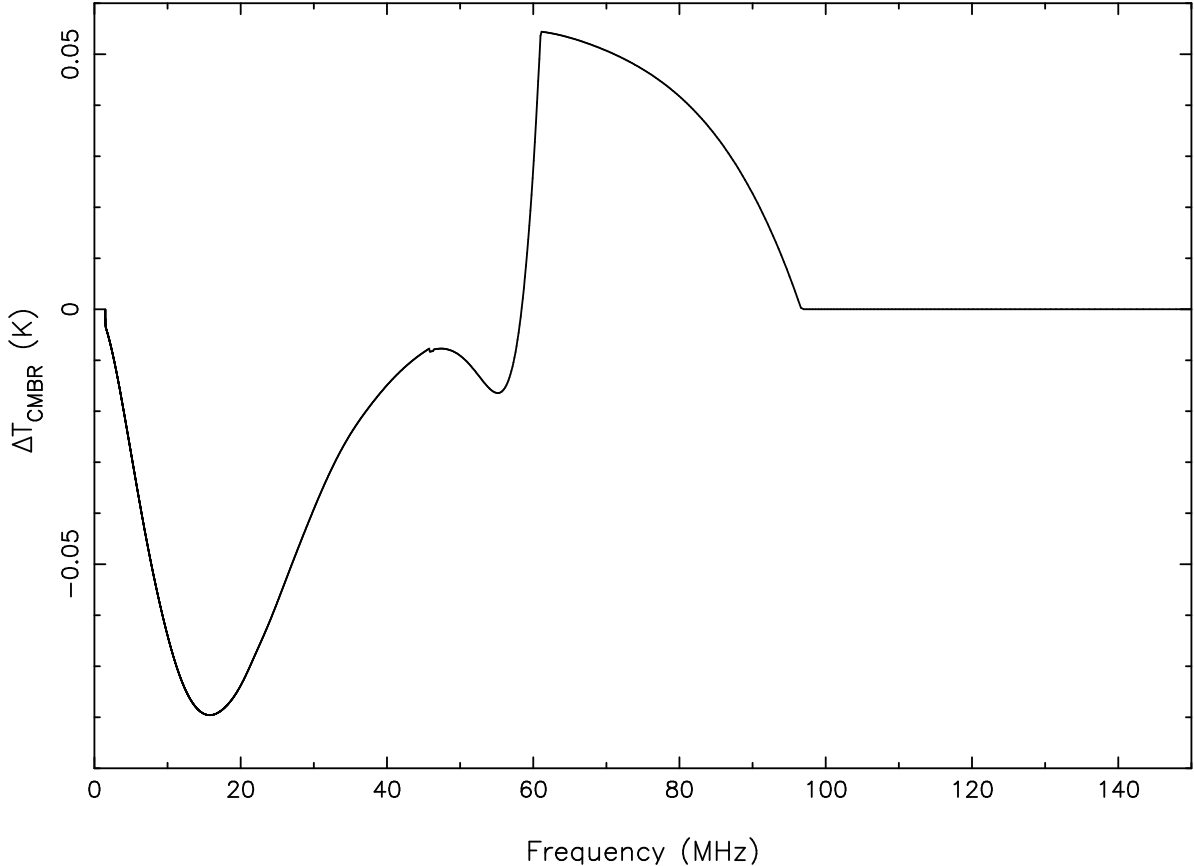}
\caption{$\Delta T_{\rm CMBR}$ (Eq.~(\ref{eqh1})) is shown 
as a function of the observed frequency for an ionization history in which
the universe becomes fully ionized at $z = 15$}
\label{fig:f6}
\end{figure}

From Eq.~(\ref{eqh1}) it is clear that HI can be observed in either 
absorption or emission against CMBR depending on whether $T_s$ is less 
than or exceeds $T_{\rm CMBR}$. At $z \simeq 1000$, $T_{\rm CMBR} = T_K$ and 
it follows from Eq.~(\ref{eqts}) that $\Delta T_{\rm CMBR} = 0$. As seen 
in Figure~1, for $z \la  100$, $T_K < T_{\rm CMBR}$ and therefore HI can 
be observed in absorption against CMBR if $y_c \ga 1$. During reionization,
$y_\alpha$ term can become important and  owing to thermalization at frequencies
close to Lyman-$\alpha$, $T_\alpha \simeq T_K$ \cite{field2}. 
The temperature of the medium can also 
exceed $T_{\rm CMBR}$  from X-ray and Lyman-$\alpha$ heating 
(for details see \cite{madau}, \cite{sethi1})
which implies that $\Delta T_{\rm CMBR} > 0$. $\Delta T_{\rm CMBR}$ approaches 
zero as the reionization is completed. 	In Figure~6, we show $\Delta T_{\rm CMBR}$ as a function of the observed frequency for 
an ionization history in which the universe becomes fully ionized at $z = 15$
(more details in next section and  \cite{sethi1}). 
If the universe re-ionized at $z \simeq 15$, then 
$\Delta T_{\rm CMBR} \simeq 0.05 \, \rm K$ at frequencies $\nu \simeq 50\hbox{--}80 \, \rm MHz$. In addition there is a signal from pre-reionization 
epoch with $\Delta T_{\rm CMBR} \simeq -0.05 \, \rm K$ at $\nu \simeq 30 \, \rm MHz$. In addition to the average signal there will be fluctuations in the 
temperature difference owing to fluctuations in HI density from primordial
density perturbations and also from the patchiness of reionization (for 
details see e.g. \cite{tozzi}). 
Currently the prospects of detecting this signal 
are being studied by using both single dish and interferometric experiments
at low radio frequencies (\cite{shaver}, see also \cite{lofar}). 

\section{Re-ionization of the Universe at High Redshift}
In the previous section we discussed various probes of the high redshift
universe and the epoch of reionization. In this section we take up the issue 
 whether it is possible to explain the observed ionization structure 
of the universe within the framework of currently-favoured $\rm \Lambda CDM$ models
of formation of structures. In these models, the observed structures in the 
universe grew from  gravitational instability of density fluctuations which 
originated during inflationary epoch in the very early universe 
(see .e.g \cite{peebles1}, \cite{padmanabhan1,padmanabhan2}). The gravitational collapse 
of these structures
might either set off star-formation  (alternatively some material might 
end up in black holes which by accreting more matter will radiate  with
harder spectrum than first star-forming galaxies, see e.g. \cite{ricotti}) which will emit 
UV light and ionize the IGM. The process of re-ionization of the 
universe is generally quite complicated and not well understood. However 
it is possible to study it  within the framework of simple 
models which might  give important clues about the details of this 
process.  Important ingredients of this problem are: (a) Halo population
at high redshift, (b) molecular and atomic cooling in Haloes, (c) Initial 
Mass Function of stars and star formation rate, (d) Escape fraction of 
UV  photons from Haloes, (e) clumpiness of the IGM. Of these the most uncertain
are (c) and (d) and have to be modelled using simple parameterized model. 

{\it Halo Population}: The number density of dark matter haloes per unit 
mass $M$ at any redshift can be obtained from the Press-Schechter method (see .e.g \cite{peebles1}, \cite{padmanabhan1,padmanabhan2}). It is given by:
\begin{equation}
{dn \over dM} = \sqrt{{2\over\pi}}{\rho_m \over M}\delta_c(z) \left | {d\sigma \over dM} \right | {1 \over \sigma^2(M)} \exp\left [-\delta_c(z)^2/(2\sigma^2(M)) \right ]
\label{eq:psme}
\end{equation}
Here $\sigma(M)$ is the mass dispersion filtered at the scale corresponding 
to mass $M$ in the linear theory; $\sigma(M)$ for length scale 
corresponding to $8 h^{-1} \, \rm  Mpc$ is $\simeq 0.9$ \cite{spergel}; $\delta_c(z) \simeq 1.7 D(0)/D(z)$, 
with $D(z)$ being the solution of growing mode in linear theory; in sCDM model
$D(0)/D(z) = (1+z)$ (for details see \cite{padmanabhan1,padmanabhan2}, \cite{peebles2}). The first baryonic 
structures to collapse would have masses exceeding the Jeans mass of the 
IGM, whose evolution  is shown in Figure~1. From Figure~1 it is seen that
the first structures might have masses $\simeq 10^4 \, \rm M_\odot$
(for non-linear extension of the concept of Jeans mass see e.g. \cite{barkana}; 
the modified  mass has values similar to the one obtained using linear theory). The  collapsed fraction of all structures can be calculated from Eq.~(\ref{eq:psme}) and is $\simeq 2 \times 10^{-3}$ at $z \ simeq 20$. However to form
stars, baryons need to cool sufficiently rapidly in the dark matter haloes
to collapse to density higher than initial  virialized density.

\begin{figure}
\epsfig{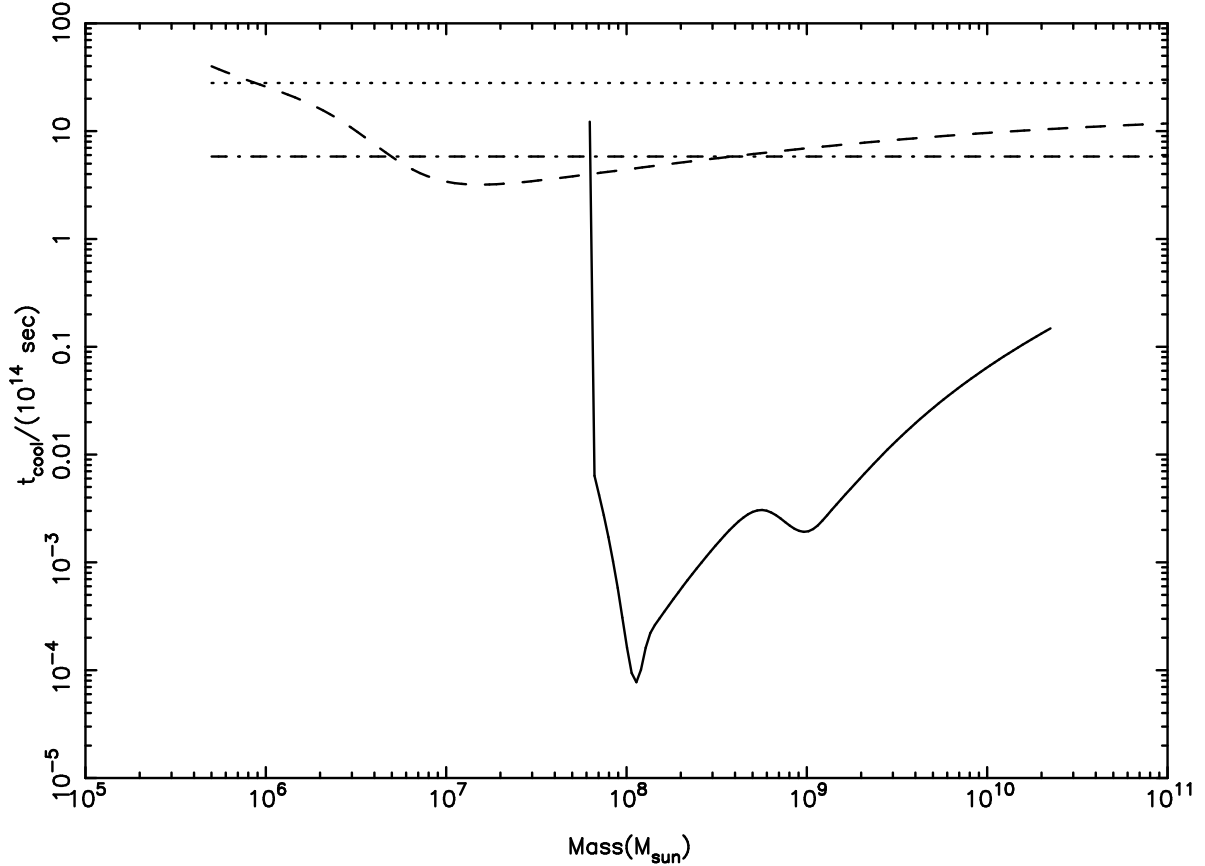}
\caption{Cooling time is shown against the mass of the virialized
halos at $z = 20$. The solid line is from atomic cooling of the 
primordial gas containing only  hydrogen
and helium. 
 The cooling processes include: line cooling from hydrogen
and helium, free-free emission, and recombination.  The dashed curve
is for cooling from molecular hydrogen for a molecular 
fraction $5 \times 10^{-4}$. The dotted line is the inverse of 
 local expansion rate
$H^{-1}(z)$ and the dot-dashed line is the dynamical  time $1/\sqrt{G\rho_m}$.}
\label{fig:f7}
\end{figure}

{\it Atomic and molecular cooling}: Assuming a spherical, top-hat collapse (see
e.g. \cite{padmanabhan1,padmanabhan2}) the density of the collapsed structure is 
$\simeq 170$ times the background density at that redshift. If a virial 
equilibrium is reached inside the halo, the baryon temperature is  
raised to the virial temperature given by (e.g. \cite{padmanabhan1,padmanabhan2}): 
\begin{equation}
T_{\rm vir} \simeq  800 \, {\rm K} \left ({M \over 10^6 M_\odot} \right )^{2/3} \left ({1+z \over 20}\right) \left ( { \Omega_m \over 0.3} \right )^{1/3}\left ( {h \over 0.7} \right )^{2/3}  \left ( { \mu  \over 1.22} \right )
\end{equation}
Here molecular weight $\mu = 1.22$ for a fully neutral halo (haloes with
masses $\la 10^8\, \rm M_\odot$ at $z \simeq 20$); $\mu = 0.57$ for a fully ionized halo. An important criterion is whether baryons can cool rapidly enough 
so that they collapse to higher densities, fragment and form stars. 
In the primordial gas, the cooling
in haloes with virial temperature $\ga 10^4$ is dominated by atomic 
hydrogen and singly-ionized helium. But the smaller haloes can only cool further by molecular hydrogen. A small fraction $\simeq 10^{-6}$ of molecular 
hydrogen is formed at $z \simeq 1000$ 
 in the IGM following recombination of the universe (see e.g. \cite{peebles1}). 
Such a small fraction doesn't suffice to cause rapid enough cooling. However
the collapse of halo  can cause  the formation of molecular hydrogen,
 resulting in a molecular fraction
$\simeq 5 \times 10^{-4}$ by the time the halo virializes
 (see e.g. \cite{tegmark}  and reference therein). In Figure~7, we
show the cooling time (defined as $t_{\rm cool} = k T_K/\dot E$, where $\dot E$ is the rate
at which the halo loses energy from various processes) for haloes of 
different masses at $z = 20$. The criterion for runaway cooling resulting 
in fragmentation and star formation is that cooling time be  less than
the dynamical time ($t_d = 1/\sqrt{G\rho}$) of the halo.  From
Figure~7 it is seen than haloes of masses $\simeq 10^6 \, \rm M_\odot$ can 
collapse to form stars at $z \simeq 20$. This result is only approximate
as the haloes will not have constant density. For 
instance, the more dense central parts
of the halo can form both more molecular hydrogen and cool more rapidly.
Bromm et al. \cite{bromm} simulated the collapse of haloes of masses $\simeq 10^{6} \rm M_\odot $ with substructure.  They concluded that the halo fragments into 
many clumps with typical masses $10^2 \hbox{--}10^3 \, \rm M_\odot$; their 
analysis suggests that this mass scale is a result of molecular hydrogen
chemistry and therefore should correspond to the masses of the first stars. 
Abel et al. \cite{abel} reached similar conclusions from their simulations.
 
\begin{figure}
\epsfig{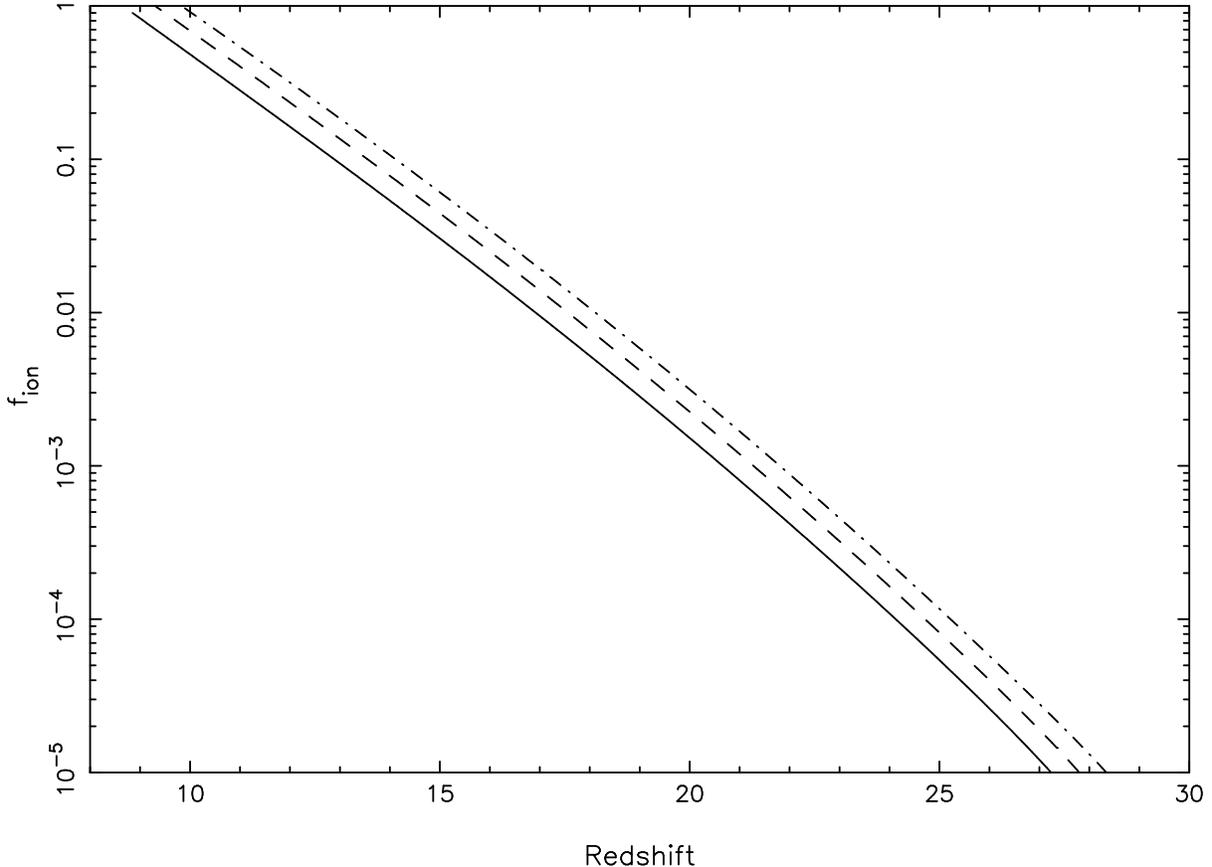}
\caption{Ionization history is shown for several values of $ N_\gamma(0)$
and $C$ (see text for details). Solid, dashed, and dot-dashed curves
are: \{$N_\gamma(0) = 10^{50} \, \rm sec^{-1}$, $C = 1$; $N_\gamma(0) = 2 \times 10^{50} \, \rm sec^{-1} $, $C = 2$; $N_\gamma(0) = 5 \times 10^{50} \, \rm sec^{-1}$, $C = 5$\}, respectively.}
\label{fig:f8}
\end{figure}

From information about the halo population and cooling arguments it is 
possible to speculate on the ionization history of the universe from
photo-ionization. The main uncertainty is the hydrogen-ionizing luminosity
(and its evolution) of each halo, which has to  be parameterized. Assuming 
that  halo of mass $M$ emits isotropically the hydrogen-ionizing luminosity
$\dot N_\gamma$ (in photons~$\rm sec^{-1}$), the radius of ionizing sphere 
around the source will satisfy the equation (Stromgren Sphere, see e.g. 
\cite{shu}, \cite{shapiro}):  
\begin{equation}
{dR \over dt} -H R={(\dot N_\gamma - 4\pi/3 R^3 \alpha_B C  n_b^2 x_{\rm \scriptscriptstyle HI})  \over (\dot N_\gamma + 4 \pi R^2 x_{\rm \scriptscriptstyle HI} n_b)}
\end{equation}
Here $C\equiv \langle n_b^2 \rangle/\langle n_b \rangle^2$ is the clumping 
factor of the IGM.
Using Eq.~(\ref{eq:psme}), the fraction of the universe that is ionized as a given redshift is (see e.g. \cite{haiman1}, \cite{sethi1}):
\begin{equation}
f_{\rm ion}(z) = {4 \pi \over 3}\int_{0}^z dz' \int dM {dn \over dM}(M,z') R^3(M,z,z')
\end{equation}
Further assuming that the luminosity of the source is $\propto M$, the 
ionized fraction can be calculated in terms of the evolution of the photon
luminosity of a single halo of some fiducial mass and the evolution of 
the clumping factor. For simplicity we assume the clumping factor to have 
a constant value between one and five, and take the luminosity evolution of 
a halo to have the form of a typical star-burst galaxy: $\dot N_\gamma(t) = \dot N_\gamma(0)\exp(-t/{10^7 \, \rm years})$. In Figure~8 we show several
ionization histories for different values of $N_\gamma(0)$ 
for  $M = 5 \times 10^7 \, \rm M_\odot$ and clumping factor $C$.  
In Figure~8 we only
plot ionization fraction upto $z \simeq 9$ at which $f_{\rm ion} \simeq 1$;
further evolution keeps the universe fully ionized upto the present.
These ionization
histories are consistent with WMAP observations. 
(If the universe becomes fully
ionized at $z \simeq 8$ if will remain ionized upto the present even in
the absence of ionizing sources 
 as the typical recombination time
$1/(\alpha_B C n_b)$ already exceeds the age of the universe at this epoch
for $C \la 2$.) It appears that $ N_\gamma(0) \simeq 10^{50}$ is required 
to ionize the universe early enough to satisfy WMAP observations. This is 
just three orders of magnitude below the photon luminosity of a typical
star-burst galaxy (see .e.g.\cite{leitherer}). 
Alternatively one can infer  that the efficiency of the first star
formation was very high (see e.g. \cite{haiman1}, \cite{chui}). There are other 
uncertainties like feedback from supernova, and photo-dissociation of 
molecular hydrogen which indicate this estimate is  a lower limit 
(see e.g. \cite{barkana}, 
 \cite{haiman1}, \cite{haiman2}, \cite{dekel}). 
Alternatively it is possible that the collapsed fraction of the universe
far exceeded the value given by $\rm \Lambda CDM$ models and was caused 
by some other physical process like tangled magnetic fields \cite{sethi2}.

Is it possible to simultaneously  satisfy the WMAP  and
the GP observations? It appears difficult to achieve it unless the 
efficiency of star formation decreased rapidly 
from $z \simeq 20$ to $z \simeq 6$ (for other possible scenarios and 
details see e.g. \cite{haiman1}, \cite{chui} 
and references therein). 

\section{Discussion}
The ionization history of the universe at high redshift as
inferred by WMAP and GP observations is quite complicated: the universe 
reionized at $z \simeq 12\hbox{--}15$;  the neutral fraction increased to 
better than one part in a thousand
 for $5.5 <z < 6.5$; and for $z \la 5$, the neutral
fraction dropped to one part in a million. Theoretical analyses
based on models of structure formation  are not
sophisticated enough to understand  this ionization history. Future observations
however might throw light on these observations. Firstly, future telescopes
might be able to detect the sources of reionization upto $z \simeq 20$
(\cite{haiman1}). Future CMBR experiment Planck also  
is sensitive enough
to disentangle the effects of complicated ionization histories \cite{kaplinghat}.

\end{document}